# Compact Three Mirror Anastigmat Space Telescope Design using 6.5m Monolithic Primary Mirror

Daewook Kim *[a,b,c], Heejoo Choi[a,c], Ewan Douglas[b]

[a]James C. Wyant College of Optical Sciences, University of Arizona, 1630 E. University Blvd., Tucson, AZ 85721, USA
[b]Department of Astronomy, University of Arizona, 933 N. Cherry Ave., Tucson, AZ 85721, USA
[c]Large Binocular Telescope Observatory, 933 N. Cherry Ave., Tucson, AZ 85721, USA

## ABSTRACT

The utilization of a 6.5m monolithic primary mirror in a compact three-mirror anastigmat (TMA) telescope design offers unprecedented capabilities to accommodate various next generation science instruments. This design enables the rapid and efficient development of a large aperture telescope without segmented mirrors while maintaining a compact overall form factor. With its exceptional photon collection area and diffraction-limited resolving power, the TMA design is ideally suited for both the ground and space active/adaptive optics concepts, which require the capture of natural guide stars within the field of view for wavefront measurement to correct for misalignments and shape deformation caused by thermal gradients. The wide field of view requirement is based on a statistical analysis of bright natural guide stars available during observation. The primary mirror clear aperture, compactness requirement, and detector pixel sizes led to the choice of TMA over simpler two-mirror solutions like Ritchey-Chretien (RC) telescopes, and the TMA design offers superior diffraction-limited performance across the entire field of view. The standard conic surfaces applied to all three mirrors (M1, M2, and M3) simplify the optical fabrication, testing, and alignment process. Additionally, the TMA design is more tolerant than RC telescopes. Stray light control is critical for UV science instrumentation, and the field stop and Lyot stop are conveniently located in the TMA design for this purpose.

## 1. INTRODUCTION

The 2020 Decadal Survey states that "A large-aperture UV/optical space telescope, however, also envisaged for addressing Pathways to Habitable Worlds, would transform this subject. The combination of 6 m-class aperture and a high-efficiency spectrograph with modern detectors would provide thousands of potential sightlines to nearby galaxies, enabling "tomographic" studies of their circumgalactic and interstellar media, as well as rich new observations of the intergalactic gas clouds along the same lines of sight." [1] The compact three-mirror anastigmat (TMA) telescope design allows a swift and effective development of a 6.5 m class telescope utilizing "non-segmented" large aperture. This telescope design will boast a vast photon collection area and achieve diffraction-limited resolving power, all thanks to its 6.5 m diameter monolithic primary mirror, which will be especially ideal for various future space telescope missions.

## 2. TELESCOPE DESIGN BOUNDARY CONDITIONS & REQUIREMENTS

### 2.1 Readily Available Primary Mirror

The compact TMA space telescope design study takes advantage of the existing readily available 6.5 m primary mirror (M1) substrate at the University of Arizona. The mirror is about 90% light-weighted and ready for the immediate computer-controlled fabrication and figuring process. This enables rapid telescope development within the target timeline. The given boundary condition from the existing mirror substrate is as below.

• Boundary Condition-1: The M1 Clear Aperture (CA) size and Radius of Curvature (RoC) are set to the as-casted mirror value. M1's maximum CA is 6.46 m and the RoC is 16.256 m.

*dkim@optics.arizona.edu

## 2.2 Launch Rocket Fairing Size Availability

A launch rocket similar to SpaceX's Starship is assumed in order to take full advantage of the newly available capabilities. [2] Thus, the fairing size is approximately given and the telescope must fit inside the envelope (considering other key structures form-factors such as primary mirror supporting structures, science instruments, and telescope baffles). This becomes the second boundary condition.

- Boundary Condition-2: The maximum Total Optical Length (TOL), which is the longest distance between two optical surfaces in the telescope design, is set to ~8 m.

It is worth noting that the compact form-factor provides good benefits for ground telescope applications, also. A compact telescope allows smaller telescope enclosure, which consumes a large portion of the overall project cost. Furthermore, a compact telescope design (e.g., Rubin Observatory) can move from one observation point to the other quickly because of its superb opto-mechanical and structural performances.

## 2.3 Telescope F/# for Commercially Available Detectors

The telescope focal plane detector (i.e., context camera sensor) needs to be a readily available commercial detector. While customized detectors may deliver tailored performance, such detectors may dramatically increase the overall project cost. Also, lead-time of a large format detector plane development is often in the order of multiple years. In order to accomplish a rapid and cost-efficient mission, an array (e.g., 13 detectors using 3 by 5 array with two missing detectors to feed the instruments through the focal plane) of commercially available Sony IMX455 detector (or similar) is chosen. The detector will be used without modifying any of its form-factor. Its pixel size is around $3.76 \times 3.76$ µm. The distance between pixel centers is also assumed to be about 3.76 µm.

Because the telescope focal plane's context camera image can be used for the wavefront sensing (e.g., phase retrieval measurement, SPGD (Stochastic Parallel Gradient Descent) optimization search, curvature sensing) the focal plane PSF (Point Spread Function) needs to be sampled higher than the minimum $2 \times 2$ pixels. About $6 \times 6$ detector pixels per PSF was assumed. Thus, Airy disk PSF diameter, which is about $2.44\lambda \cdot F/\#$, needs to be about 22.56 µm (= 6 pixels × 3.76 µm/interpixel). Assuming a design wavelength of $\lambda = 0.6$ µm, the STP telescope's F/# needs to be about 15 (= 22.56 / 2.44 / 0.6). This sets the third boundary condition of the telescope design as below.

- Boundary Condition-3: In order to match the available detector pixel size and PSF sampling, the STP telescope's F/# needs to be about 15.

## 2.4 Telescope FoV Requirement

The telescope concept necessitates capturing natural guide stars within the field of view (FoV) to enable wavefront measurements for active telescope alignment and M1 mirror shape correction. Since there are no laser-guide stars in space, the FoV requirement was determined using statistics of the bright natural guide stars visible in the sky during observations. This requirement is crucial for achieving diffraction-limited performance of the telescope in the challenging space environment, where thermal gradients can potentially lead to misalignments and shape deformations of the 6.5 m diameter M1 mirror.

- Optical Design Requirement-1: The required FoV coverage area for natural guide star acquisition is about 6 mrad$^2$. This is taking account of various factors such as statistics of the stars/brightness and the empty "null" FoV areas due to the dead gaps between each Sony detector units (with boundary packaging areas around the active detector zone).

## 3. TWO-MIRROR VS. THREE-MIRROR DESIGN TRADE-OFF

Two-mirror telescope design such as Ritchey-Chretien (RC) telescope provides a simpler telescope configuration compared to three-mirror design solutions such as TMA (Three Mirror Anastigmat). Also, on-axis TMA often requires M4 in order to fold the beam towards the science instruments behind M1. Thus, a flat M4 is added to a typical TMA design such as the James Webb Space Telescope. [3]

The trade-off study between two- and three- mirror telescopes has been performed and reported by optical designers. Feinberg et al. states "In general, each powered mirror provides correction for a single third-order aberration. A single-mirror system can correct spherical aberration (i.e., Off-Axis Parabola design), but coma and astigmatism remain, both of which cause blur off-axis. A two-mirror system, on the other hand, can be designed to be free of both spherical aberration and coma [i.e., Ritchey-Chretien design (RC)], leaving only astigmatism as a blurring influence on the image quality. The two-mirror design was adopted for the HST (Hubble Space Telescope) where proper correction allows for reasonable imaging across several arcminutes of field of view. A three-mirror system provides correction of third-order spherical aberration, coma, and astigmatism, allowing for fields of view of 20 arcmin (e.g., JWST (James Webb Space Telescope)) and up to several degrees [e.g., the Wide-Field Infrared Survey Telescope (WFIRST)]. These designs are referred to as "Korsch three-mirror anastigmats" (TMAs), and are becoming more common as science demands larger fields of view for consideration." [4]

Based on the design benchmark study for the 1-m aperture F/10 telescope case [4], the equivalent plot for the 6.5-m aperture F/10 telescope case will be about 6.5 times worse (i.e. larger) in wavefront error. While there is room for further numerical optimizations and the compact TMA space telescope's F/15 system may allow slightly better wavefront performance compared to the F/10 (i.e., optically faster system) benchmark, the 6.5× wavefront error degradation confirms that the FoV of 2-mirror RC designs cannot provide the required FoV coverage of 6 mrad$^2$. For instance, even the "center field optimize" RC case [4], which provides diffraction-limited performance around the center field, begins to suffer from rapid increase of the wavefront error beyond ~0.75 mrad FoV (or ~0.5 mrad$^2$ for a square FoV coverage). This is an order of magnitude smaller than the required 6 mrad$^2$ FoV coverage. Thus, the compact TMA space telescope chose the TMA design solution.

As another important sidenote, Feinberg et al. states "In general, three-mirror systems are also more tolerant of misalignment since they are corrected for the third-order image blur aberrations. When the secondary mirror (SM) is decentered by 10 μm, for example, the two-mirror system degrades by ~24 nm RMS wavefront error (WFE), whereas the three-mirror system degrades by only ~8 nm RMS WFE." [4]

## 4. COMPACT TMA SPACE TELESCOPE DESIGN

### 4.1 Compact TMA Space Telescope Design Overview

The compact TMA space telescope design specification is summarized in Table 1. This meets the three Boundary Conditions (BCs) and Optical Design Requirement-1 stated in Section 2.

Table 1. Compact TMA Space Telescope Optical Design Specifications

| Optical Design Parameter | Nominal Value | Note |
|---|---|---|
| Telescope Design Type | TMA | Common Optical Axis Three Mirror Anastigmat (TMA). |
| M1 CA Outer Diameter | 6.42 m | CA extends 40 mm from the outer edge of polished surface. |
| M1 CA Inner Diameter | 1.38 m | CA extends 34 mm from the inner edge of polished surface. |
| M1 RoC | 16.256 m | |
| M1 Conic Constant | -0.995357 | |
| M2 RoC | 1.660575 m | 780 mm mechanical diameter. 760 mm CA diameter. |
| M2 Conic Constant | -1.566503 | |
| M3 RoC | -1.830517 m | 580 × 380 mm mechanical diameter. 560 × 360 mm CA. |
| M3 Conic Constant | -0.718069 | |
| M3 Off-axis Distance | 240 mm | Distance from the parent optical axis to the center of CA. |
| M4 RoC | Inf (i.e., flat) | 150 mm mechanical diameter. 130 mm CA diameter. |
| Total Optical Length (TOL) | ~8 m | The actual thickness of mirrors may change this value. |
| F/# | 15 | |
| Nominal Design Wavelength | 650 nm | Achromatic all reflective design. |
| Field of View | +/- 0.115×0.043° | Equivalent to 4 × 1.5 mrad full FoV coverage (i.e., 6 mrad$^2$) |
| Field Bias | 0.17° | Center field bias for common optical axis TMA design. |
| Folding Mirror M4 Shape | Flat | This is ~100 mm diameter folding flat following the TMA. |
| Final Focal Plane Size | ~400 × 155 mm | |

The compact TMA space telescope optical layout is shown in Figure 1. It shows the four mirrors (M1 – 4) including the folding flat M4. M1, M2, and M3 are standard conic surfaces without using any higher order aspheric terms (i.e., no freeform surfaces). This greatly simplifies the optical fabrication, testing, and alignment of the telescope system. Also, the field bias of 0.17° is providing a good mechanical margin around the M4 to avoid any vignetting.

The M3 and M1 are almost in a co-plane so it allows sufficient space right behind the M1. This is directly related to the key requirement set by the TOL Boundary Condition-2 in Section 2.2. This is a deviation from typical TMA designs which often locate M3 behind M1 just like the JWST. [3]

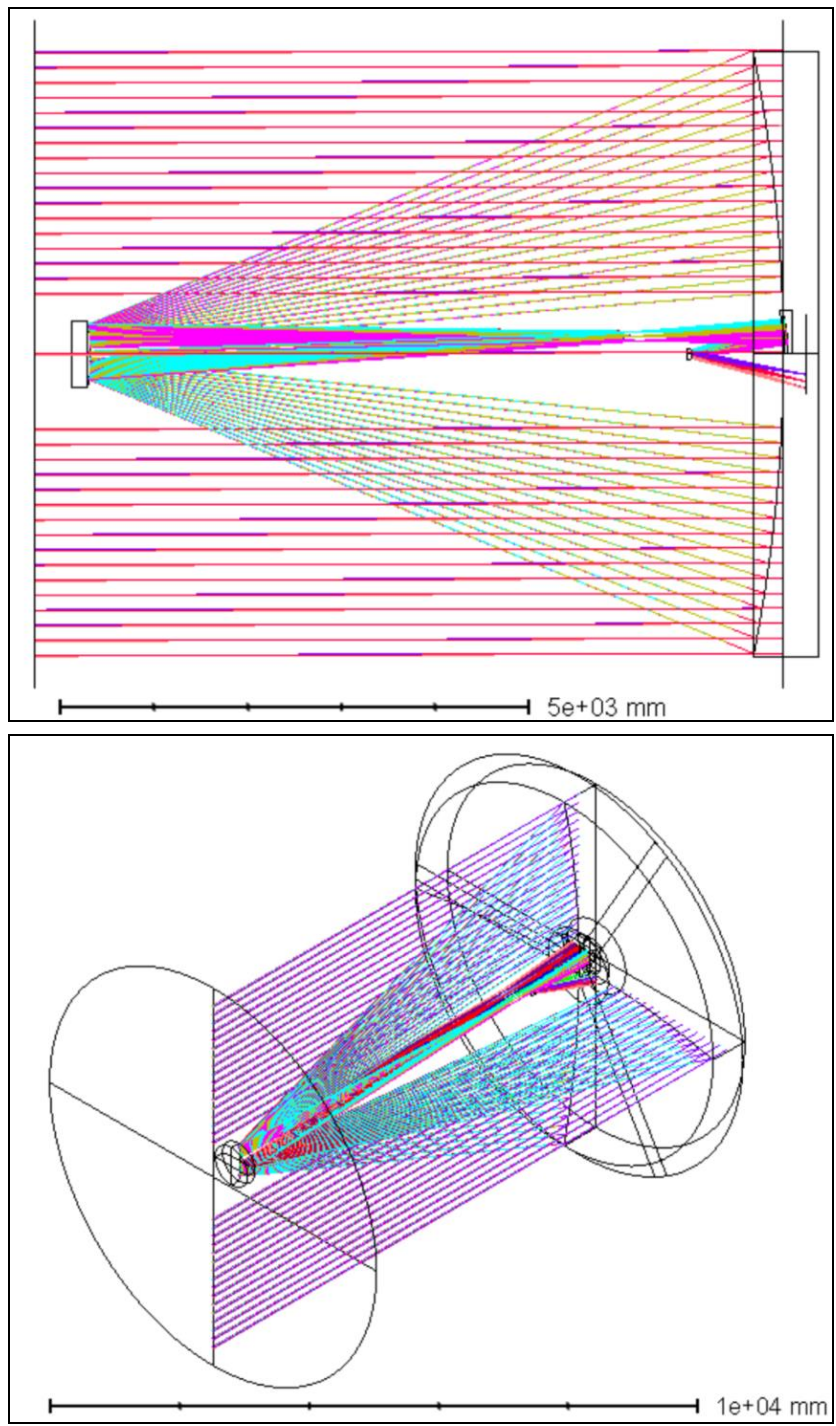

Figure 1. Compact TMA space telescope optical layout (top) showing four mirrors, M1, M2, M3, and M4. Different colors represent the FoVs of the telescope. Also, the iso-view (bottom) shows the notional shadow of the secondary mirror (M2) spider structure's shadow.

The diffraction-limited nominal design performance of the compact TMA space telescope is presented in Figure 2. The RMS radius of the spots (e.g., ~1 – 4 μm) across the full FoV are much smaller than the Airy Radius (i.e., 11.97 μm at 650 nm design wavelength).

### 4.2 Intermediate Stops for Stray Light Control and Exit Pupil for Deformable Mirror

The stray light control from the optical surface contamination/scattering or opto-mechanical structures illuminated by the out-of-field illumination (e.g., sun) is essential for the space telescopes where the sun or other bright objects may locate at a certain angle with respect to the space telescope. The STP TMA design provides an intermediate focal plane between the M2 and M3 as a Field Stop mask location. This can efficiently block the out of field illuminations. Also, the M4 location is the exit pupil location of the TMA (defined by M1, M2, and M3). Thus, the flat M4 location can be a natural Lyot Stop location to block any diffraction or edge scatterings from the M1.

As another optional-yet-powerful design feature, the M4 can be replaced with a potential DM (Deformable Mirror) to correct any M1 surface shape errors as it is the optically conjugate pupil plane.

These intermediate conjugate planes are added benefits of the TMA design compared to a two-mirror design.

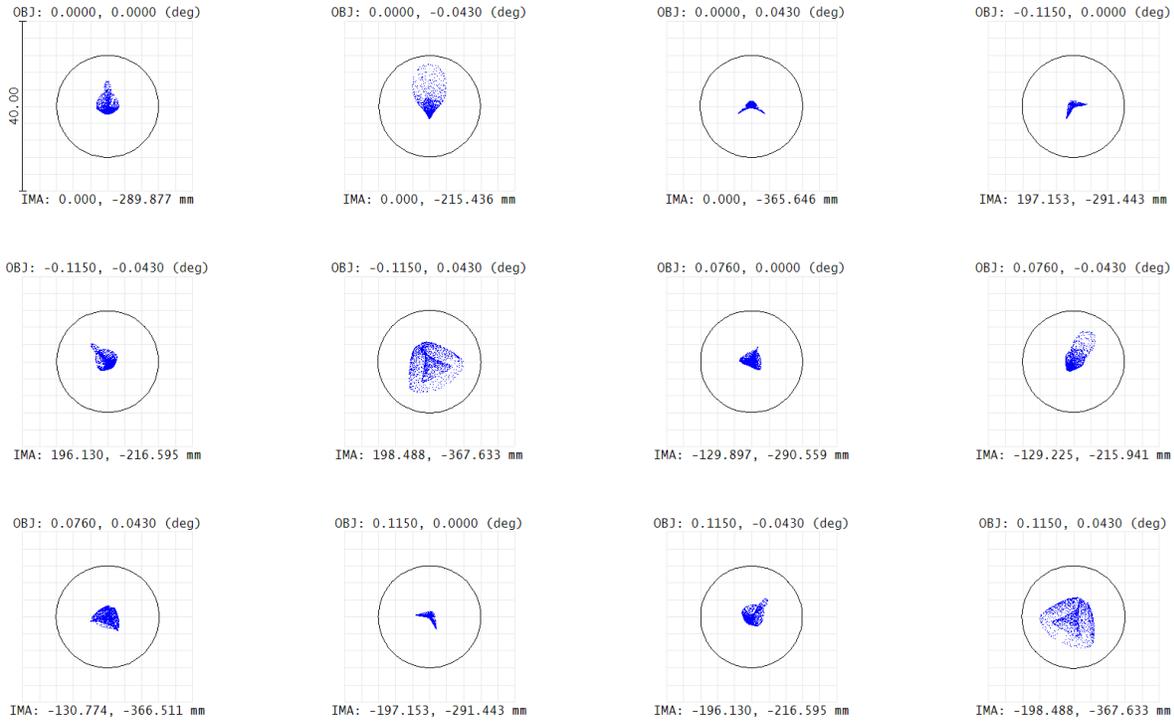

Figure 2. Compact TMA space telescope nominal spot diagram. The black circle represents the Airy disk at 650 nm design wavelength. They are all well within the diffraction-limited Airy disk circles across the entire FoV. (This simulation utilizes the full 6.46 m diameter M1 surface aperture (instead of applying the CA listed in Table 1). The spot diagram with CA will be better than the diagrams shown in this figure.

# 5. CONCLUSION

Based on the given boundary conditions (M1 dimensions, M1 RoC, Launch Vehicle Fairing Size, Detector Pixel Dimensions), the overall telescope specification such F/# and TOL was set. The FoV requirement was set by the need for natural guide stars during the observations. These conditions lead to the design choice towards TMA over simpler two-mirror solutions such as RC telescope. The nominal performance of the compact TMA design provides superb diffraction-limited performance over the entire FoV. All M1, M2, and M3 are standard conic surfaces without any higher order aspheric terms, so the optical fabrication, testing, and alignment are more straightforward. Also, TMA design is more robust than an RC telescope in terms of tolerance while TMA still has more optical surfaces to integrate and align than the two-mirror design. M4 is a relatively small ~100 mm diameter flat mirror. Both Field Stop and Lyot Stop are conveniently located for stray light control.

# ACKNOWLEDGEMENTS

Portions of this research were supported by funding from the Technology Research Initiative Fund (TRIF) of the Arizona Board of Regents and by generous anonymous philanthropic donations to the Steward Observatory of the College of Science at the University of Arizona.